\documentclass[aps,prl,reprint,
superscriptaddress,
amsmath,
amssymb,
longbibliography,
]{revtex4-2}
\usepackage{comment}
\usepackage{color}
\usepackage{graphicx}
\usepackage{subfigure}
\usepackage{dcolumn}
\usepackage{bm}
\usepackage{multirow}
\usepackage{amsmath}
 
\usepackage{braket}
\usepackage{soul,xcolor}
\usepackage[normalem]{ulem} 
\usepackage{siunitx} 

\usepackage{diagbox, eqparbox, hhline}
\usepackage{hyperref}
\usepackage{tikz-cd}
\tikzcdset{every label/.append style = {font = \normalsize}}

\DeclareMathAlphabet{\mathpzc}{OT1}{pzc}{m}{it}

\begin{document}


\title{ 
Emergent Self-propulsion of Skyrmionic Matter in Synthetic Antiferromagnets \\
\small{\color{blue} {\it Published in:} \href{https://journals.aps.org/prl/abstract/10.1103/c2y9-3cc9}{{\it Phys. Rev. Lett.} {\bf 135}, 086701 (2025)}}}

\author{Cl\'ecio C. de Souza Silva}
\email{clecio.cssilva@ufpe.br}
\affiliation{
Departamento de Física, Centro de Ciências Exatas e da Natureza, Universidade Federal de Pernambuco, Recife--PE, 50670-901, Brazil
}
\author{Matheus V. Correia}
\affiliation{
Departamento de Física, Centro de Ciências Exatas e da Natureza, Universidade Federal de Pernambuco, Recife--PE, 50670-901, Brazil
}
\author{J. C. Piña Velásquez}
\affiliation{
Núcleo de Tecnologia, Centro Acadêmico do Agreste, Universidade Federal de Pernambuco, Avenida Marielle Franco, Caruaru-PE, 55014-900, Brazil
}

\begin{abstract}

Self-propulsion plays a crucial role in biological processes and nanorobotics, enabling small systems to move autonomously in noisy environments. Here, we theoretically demonstrate that a bound skyrmion-skyrmion pair in a synthetic antiferromagnetic bilayer can function as a self-propelled topological object, reaching speeds of up to a hundred million body lengths per second—far exceeding those of any known synthetic or biological self-propelled particles. The propulsion mechanism is triggered by the excitation of back-and-forth relative motion of the skyrmions, which generates nonreciprocal gyrotropic forces, driving the skyrmion pair in a direction perpendicular to their bond. Remarkably, thermal noise induces spontaneous reorientations of the pair and momentary reversals of the propulsion, mimicking behaviors observed in motile bacteria and microalgae. 

\end{abstract}

\date{\today}

\maketitle

The challenges of swimming in highly dissipative and noisy environments have driven the evolution of ingenious yet simple locomotion strategies in microorganisms. For example, bacteria and microalgae use nonreciprocal shape deformations, driven by flexible flagella, cilia, or body modulations, and navigation strategies, like run-and-tumble and run-and-reverse motion, to harness environmental noise and explore complex surroundings efficiently~\cite{Berg1972,purcell1977,brennen1977fluid,Blackburn1998,Xie2011}. These biological self-propulsion strategies have inspired the design of synthetic microswimmers for micron-scale applications that demand minimal supervision, such as targeted drug delivery~\cite{palagi2018bioinspired,elnaggar2024state}. Furthermore, active matter systems composed of self-propelled units serve as key models for exploring non-equilibrium physics and emergent phenomena~\cite{Bechinger2016_RMP,Bowick2022}, and form the basis for a class of reconfigurable smart materials~\cite{baconnier2022selective,wang2024robo}.

Prompted by these advances, a natural question arises: can similar principles be applied to create a spintronic version of self-propelled quasi-particles? In other words, can localized spin excitations be made to ``swim'' autonomously within magnetic materials? Such active spintronic matter could unlock new pathways for information transport and processing while enabling new functionalities in spintronic devices. Here, we explore this possibility by considering magnetic skyrmions—topologically protected whirlpools of spins that naturally emerge in chiral magnets and multilayers~\cite{roessler2006spontaneous,muhlbauer2009skyrmion,yu2010real,romming2013writing,nagaosa2013review,bogdanov2020review}. The nanoscopic size and highly efficient current-driven motion of skyrmions have sparked intense interest for both fundamental research and applications in spintronics and quantum computing~\cite{finocchio2016review,Fert2017review,zhang2020review,psaroudaki2021SkQuBit,zheng2021braids,Neto2022,du2022,psaroudaki2023SkQuBitReview,qin2023spintronic}. Yet, in conventional settings, skyrmions are treated as \emph{passive} particles~\cite{Kong2013,Lin2014,mochizuki2014thermally,NagaosaStochastic2014,zhang2018manipulation,Zhao2020,Weissenhoffer2021,Raimondo2022,Souza2024}, moved by external forces like spin currents, thermal gradients, or magnons, so their potential to move autonomously remains largely unexplored. 

\begin{figure}[b]
\includegraphics[width=\linewidth]{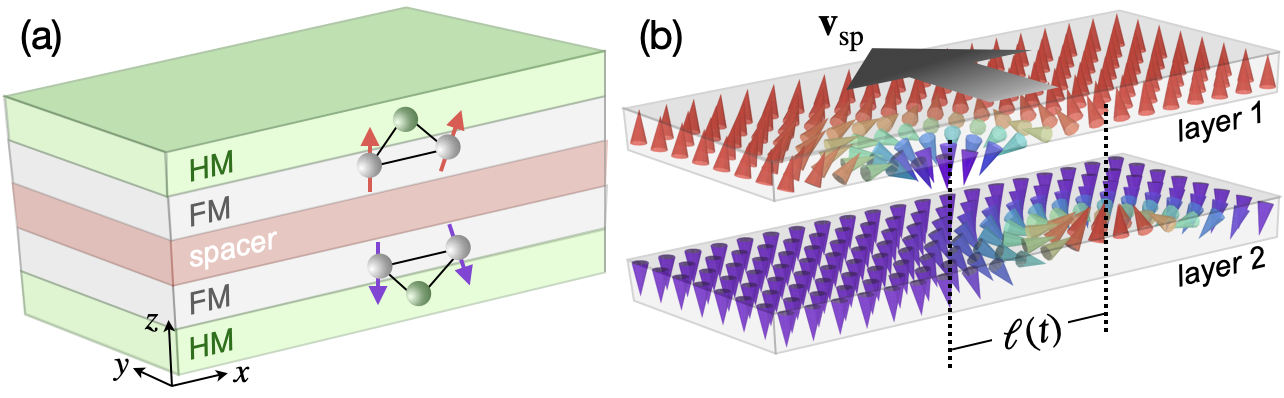}
\caption{(a) In a mirror-symmetric stack of two ferromagnet--heavy-metal (FM-HM) bilayers coupled antiferromagnetically via a non-magnetic spacer, the HM ions mediate a chiral interaction between neighboring spins in the adjacent FM layer, yielding opposite chiralities across the stack~\cite{moriya1960}. (b) Cross-sectional view of the spin configuration for two non-coaxially bound skyrmions. Self-propulsion, $\bm{v}_\text{sp}$, emerges when the bond length $\ell$ is forced to oscillate in time (see text).
}
\label{fig.Config}
\end{figure}

In this Letter, we demonstrate that bound skyrmion pairs in a synthetic antiferromagnet (SAF) can be excited into distinct ``swimming'' cycles, enabling them to move autonomously within the magnetic material as an \emph{active} nanoscopic spin texture. 
SAF systems made of two chiral ferromagnetic (FM) layers coupled antiferromagnetically can host skyrmions of opposite topological charges that form a bound state~\cite{zhang2016magnetic,KoshibaeNagaosa_2017,dohi2019formation,SK_Nucleation_ThermalEffect,Matheus2024} and display ultrafast motion under spin-polarized currents~\cite{HighSpeedSK_Dynamics_SAF}. Typically, both FM layers have the same chirality, so left-to-right spin twists across both skyrmions occur in the same direction, leading to a strongly bound coaxial state. Here we consider a variant where the FM layers have opposite chiralities [Fig.~\ref{fig.Config}-(a)]. 
The skyrmions in such a material still carry opposite topological charges, but their left-to-right spin twists reverse, making one skyrmion resemble the mirror image of the other. This means that instead of being antiparallel throughout, the magnetization at their domain walls becomes locally parallel, favoring noncoaxial binding~\cite{Matheus2024}. This arrangement, shown in Fig.~\ref{fig.Config}-(b), breaks rotational symmetry, making the skyrmion-skyrmion separation a relevant degree of freedom. 

\begin{figure*}[t!]
\includegraphics[width=\linewidth]{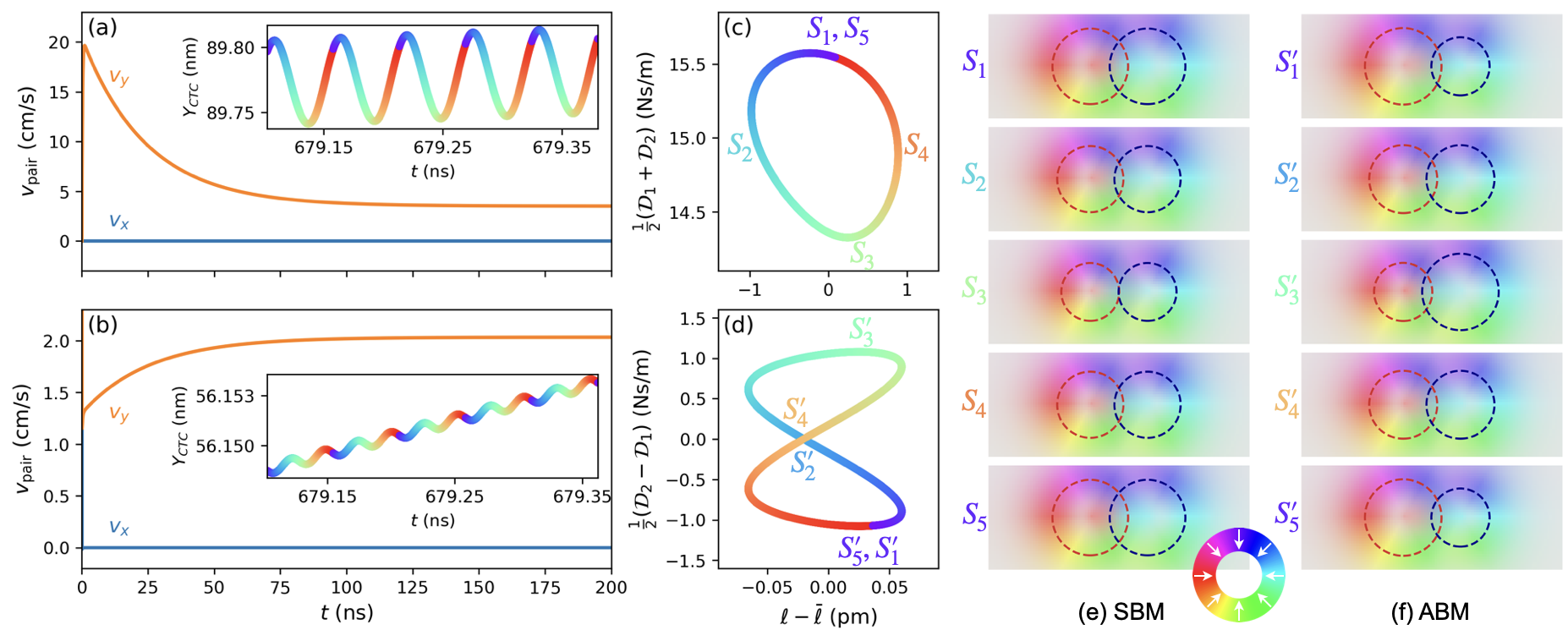}
\caption{(a) Time evolution of the skyrmion pair velocity in the directions parallel ($v_{x}$) and perpendicular ($v_y$) to the bond under SBM excitation via parametric modulation of the anisotropy coefficient at amplitude $\Delta K/K_0=0.005$ (with $K_0=0.6$ MJ/m$^3$ and $A_\text{int}=0.02$ mJ/m$^2$) and frequency 18 GHz, for Gilbert damping $\alpha=0.02$. The pair was initially equilibrated with the excitation turned off, resulting in a bond length $\ell=10.98$ nm. (b) Same as (a), but under ABM excitation via an oscillating magnetic field of amplitude 4 mT and frequency 19.24 GHz. The insets in (a) and (b) show the evolution of the perpendicular component of the center of topological charge (CTC) along 5 excitation cycles in steady state. The color cycle (blue to red) maps the excitation phase within each period. (c, d) Steady-state swimming cycles in the plane defined by the bond deformation, $\ell-\bar{\ell}$, and skyrmion shape deformation, represented by the reduced dissipation coefficients $\frac{1}{2}({\mathcal{D}}_2\pm\mathcal{D}_1)$, calculated from the micromagnetics data using Eqs.~\eqref{eq.bondlength} and \eqref{eq.D} for respectively SBM ($\bar{\ell}=10.98$ nm) and ABM ($\bar{\ell}=10.39$ nm). (e, f) Close-up view of the skyrmion pair at representative swimming stages of the (e) SBM and (f) ABM cycles. The circles depict the domain wall ($m_z=0$) of the top (red) and bottom (blue) skyrmions. The in-plane magnetization of both layers is shown by overlaying their maps with partial transparency; directions are indicated by the color wheel.
}
\label{fig.SeriesMechanism}
\end{figure*}
\bigskip
\emph{Zero-temperature self-propulsion}---We first characterize the equilibrium properties of the noncoaxial skyrmion pair by numerically solving the Landau-Lifshitz-Gilbert (LLG) equation without any external excitation. To be specific, we fixed the thicknesses of the FM layers, $d_1=d_2=0.4$ nm, intralayer exchange stiffnesses, $A_1=A_2=15$ pJ/m, Dzyaloshinskii-Moriya constants, $D_1=-D_2=3.05$ mJ/m$^2$, and saturation magnetizations, $M_{s1}=M_{s2}=0.58$ MA/m. We investigated the stability of a skyrmion-skyrmion pair as a function of uniaxial anisotropy $K_0$, out-of-plane magnetic field $B_z$, and interlayer exchange coupling $A_\text{int}$. This analysis revealed that the equilibrium pair separation (bond length) varies with $K_0$ and $B_z$ while keeping other parameters fixed (see the End Matter, Appendix~\hyperlink{app.Numerics}{A} for more details), suggesting that applying an ac electric or magnetic field perpendicular to the film can excite the longitudinal vibration of the bond inducing relative motion between the skyrmions. 

An ac electric field $\bm{E}(t)=\hat{\bm{z}}E(t)$ induces an effective anisotropy $K_\text{eff}(t)=K_0+\Delta K\sin\omega t$, where $\Delta K$ is proportional to the field amplitude~\cite{maruyama2009large,upadhyaya2015electric,yuan2019wiggling}. Because the anisotropy energy is invariant under mirror reflection, this modulation causes both skyrmions to expand and contract synchronously at a frequency $\omega$; that is, $\bm{E}(t)$ excites the symmetric breathing mode (SBM) of the skyrmions. In contrast, an ac magnetic field, $\bm{B}(t)=\hat{\bm{z}} B_0\sin\omega t$, breaks inversion symmetry and excites the antisymmetric breathing mode (ABM): under positive $B_z$, the top skyrmion contracts while the bottom one expands, and vice-versa for negative $B_z$~\cite{lonsky2020coupled}. Both breathing modes are accompanied by the longitudinal vibration of the bond at a frequency $\omega$, for SBM, and $2\omega$, for ABM. The higher frequency in the latter case arises because, by symmetry, the separation between skyrmions depends only on the magnitude of $\bm{B}$. 

\begin{figure}[t!]
\includegraphics[width=\linewidth]{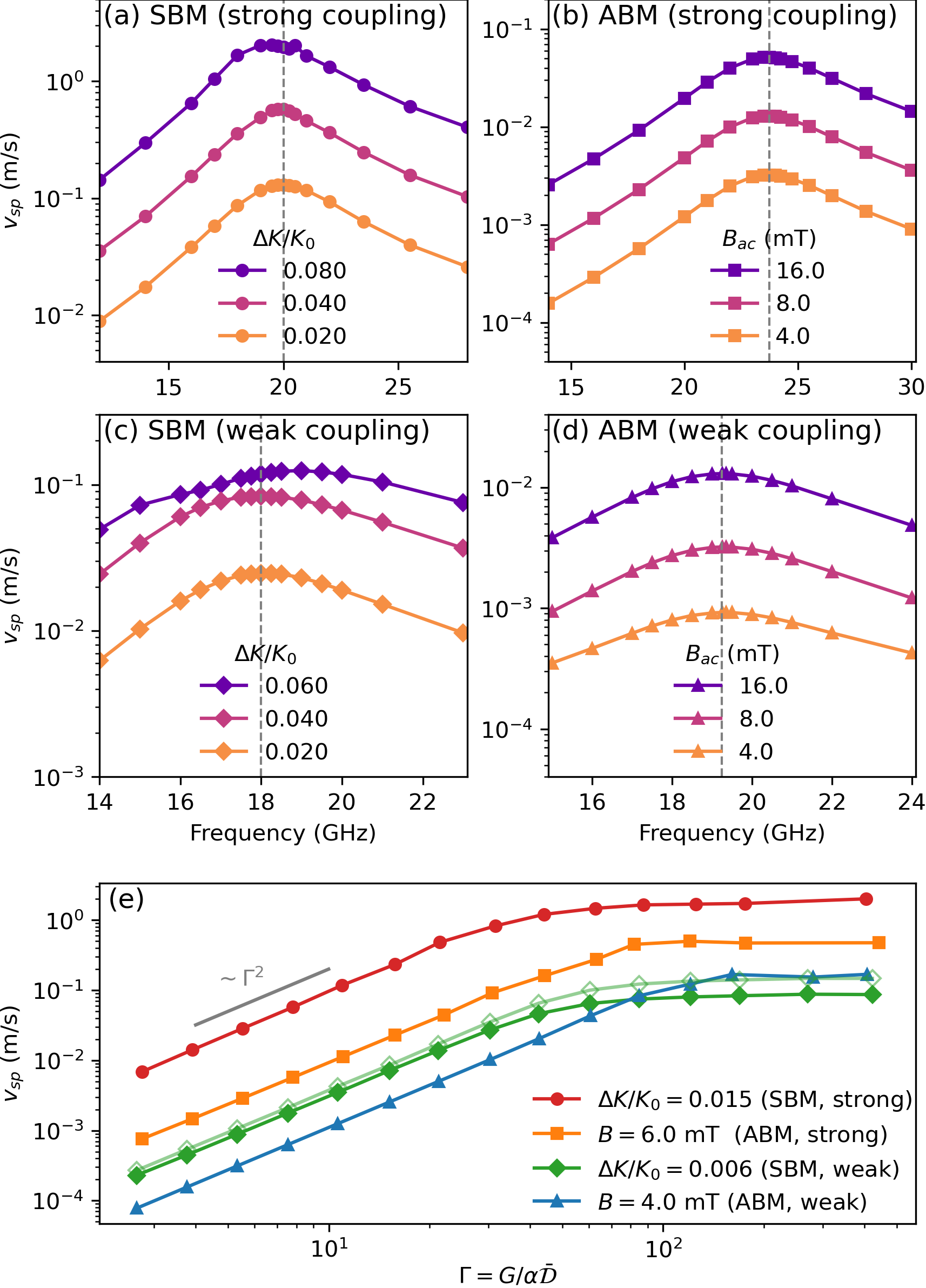}
\caption{(a, b) Resonant behavior of the self-propulsion speed ($v_\text{sp}$) of the skyrmion pair under (a) SBM, with different ac anisotropy amplitudes, and (b) ABM, with different ac magnetic field amplitudes, for $\alpha=0.1$ and strong interlayer coupling. Dashed lines indicate the resonance frequency of the corresponding mode. (c, d) Same as (a, b), but for weak coupling. (e) $v_\text{sp}$ at resonance as a function of gyrocoupling-to-friction ratio $\Gamma$ for a few selected situations. The gray line indicates $\Gamma^2$ behavior. The open symbols in (e) depict estimated $v_\text{sp}$ under weak-coupling SBM using the $\ell$-$\mathcal{D}$ integral, Eq.~\eqref{eq.vsp}. 
}
\label{fig.ResonantSelf-propulsion}
\end{figure}

Fig.~\ref{fig.SeriesMechanism} (a) and (b) shows that, under both SBM and ABM excitations, the skyrmion pair exhibits a systematic drift perpendicular to the bond, following a short transient. While the pair undergoes cyclic shape deformations, see panels (c)-(f), its center of topological charge oscillates back and forth in a direction strictly perpendicular to the bond (insets), irrespective of the bond's orientation. Crucially, this oscillatory motion is nonreciprocal, resulting in a net drift of the pair. Notice that the direction of motion is not determined by the external field, which merely supplies energy, but is instead dictated by the internal configuration and dynamics of the pair. This unambiguously demonstrates the defining hallmark of self-propelled systems: converting an energy input into autonomous motion along a direction dictated by internal degrees of freedom. Remarkably, as shown in Fig.~\ref{fig.ResonantSelf-propulsion} (a)-(d), the self-propulsion speed, $v_\text{sp}$, peaks at the resonance frequency of the corresponding breathing mode (see also Fig. S4 in the Supplemental Material~\cite{footnote_SM} and spectral analysis in End Matter, Appendix \hyperlink{app.NormalModes}{B}), reaching up to $\sim0.1$ m/s, for weak coupling ($A_\text{int}=0.02$ mJ/m$^2$, $K_0=0.6$ MJ/m$^3$), and $\sim1$ m/s, for strong coupling ($A_\text{int}=0.12$ mJ/m$^2$, $K_0=0.35$ MJ/m$^3$). Since the size of the skyrmion pair is $\sim10$ nm, this corresponds to a relative propulsion speed of up to a hundred million body lengths per second, much faster than known biological and artificial swimmers~\cite{Bechinger2016_RMP,jang2023}.

\emph{Swimming Mechanism}---At low Reynolds numbers, the propulsion speed of microorganisms and microrobots can often be expressed as an integral over their cyclic shape deformations~\cite{purcell1977}. Similarly, the self-propulsion speed of a bound skyrmion pair can be written as a closed integral in its shape manifold (see End Matter, Appendix~\hyperlink{app.Thiele}{C}). In the case of SBM under weak interlayer coupling, the skyrmions essentially keep their circular shape, reducing the shape manifold to only two internal degrees of freedom: the bond length, 
\begin{equation}\label{eq.bondlength}
  \ell=|\bm{R}_1-\bm{R}_2|,\quad\bm{R}_i = \frac{1}{Q_i}\int d^2r\,\bm{r}\varrho_i(\bm{r}),
\end{equation}
where $\varrho_i=\frac{1}{4\pi}\bm{m}_i\cdot(\partial_x\bm{m}_i \times \partial_y\bm{m}_i)$ is the topological charge density in layer $i$, $\bm{m}_i(\bm{r})$ its normalized magnetization, and $Q_i=\int d^2r\,\varrho_i$ its total topological charge, with $Q_2=-Q_1\equiv Q$ (here, the bottom skyrmion points up, so $Q=1$); and the effective dissipation coefficient of the skyrmions, 
\begin{equation}\label{eq.D}
\mathcal{D}_i\equiv\sqrt{\det{\bm{\mathcal{D}}_i}}, \quad [\bm{\mathcal{D}}_i]_{\mu\nu} = \frac{M_sd}{\gamma}\int d^2r\,\partial_\mu\bm{m}_i \cdot \partial_\nu\bm{m}_i,
\end{equation}
where $\mu,\nu=x,y$ and $\gamma$ is the gyromagnetic factor. Since the dissipation tensor $\bm{\mathcal{D}}_i$ depends only on the skyrmion shape and size, and both skyrmions are identical under SBM, $\bm{\mathcal{D}}_1=\bm{\mathcal{D}}_2$ at all times. Under weak interlayer coupling, we observe $\mathcal{D}_{xx}\simeq\mathcal{D}_{yy}\simeq\mathcal{D}$ and $\mathcal{D}_{xy}\simeq0$, indicating an essentially circular skyrmion profile (see Supplemental Material~\cite{footnote_SM}). In this regime, $\mathcal{D}$ scales directly with the skyrmion radius. Therefore, $\ell$ and $\mathcal{D}$ are the main variables determining the internal configuration of the skyrmion pair, enabling a simplified yet comprehensive visualization of the swimming mechanism.

For isotropic breathing, the translation of each skyrmion satisfies the force balance equation (see End Matter, Appendix~\hyperlink{app.Thiele}{C})~\cite{KoshibaeNagaosa_2017,Matheus2024}:
\begin{equation}\label{eq.thiele}
    \bm{G}_i\times\bm{v}_i  + \alpha{\mathcal{D}}_i\bm{v}_i = \bm{F}_i,
\end{equation}
where $\bm{v}_i=d\bm{R}_i/dt$ is the skyrmion velocity, $\bm{G}_i=4\pi(M_sd/\gamma) Q_i\hat{\bm{z}}$ is the gyrocoupling vector, $\alpha$ is the dimensionless Gilbert damping constant, and $\bm{F}_i$ is the binding force. Breaking Eq.~\eqref{eq.thiele} into components parallel ($\parallel$) and perpendicular ($\perp$) to the bond yields $v_{i\perp}/v_{i\parallel}=-G_i/\alpha\mathcal{D}$ for each skyrmion. In the SBM regime, the skyrmions mirror each other's dynamics ($v_{1\perp}=v_{2\perp}\equiv v_\perp$ and $v_{1\parallel}=-v_{2\parallel}\equiv v_\parallel$), so the pair moves strictly perpendicular to the bond, $\bm{v}_\text{pair}=(\bm{v}_1+\bm{v}_2)/2=v_\perp\hat{e}_\perp$, while $d\ell/dt = -2v_{\parallel}$. The self-propulsion speed can then be obtained by integrating $v_\perp$ over one cycle of the excitation, 
\begin{equation}\label{eq.vsp}
    {v}_\text{sp} = \bar{v}_\perp = -\frac{\omega}{4\pi}\oint\frac{G}{\alpha\mathcal{D}}d\ell.
\end{equation}
As shown in Eq.~\eqref{eq.vsp}, two elements are essential for self-propulsion. First, the coupled magnetic textures must be topologically nontrivial ($Q\neq0$). Second, the swimming cycle needs to be nonreciprocal. This nonreciprocity arises because, under SBM, $\ell$ and $\mathcal{D}$ oscillate at the same frequency $\omega$ but with different phase lags~\cite{footnote_PhaseLag}, resulting in a nonreciprocal closed cycle in $\ell$-$\mathcal{D}$ space.

Fig.~\ref{fig.SeriesMechanism} (c) and (e) show that under SBM the skyrmion pair indeed traces a closed path in the shape plane while undergoing a net displacement of its center. Furthermore, as illustrated in Fig.~\ref{fig.ResonantSelf-propulsion} (e), near the resonance frequency of the weak coupling SBM regime (18 GHz), the propulsion speed estimated from  Eq.~\eqref{eq.vsp} agrees remarkably well with that obtained directly from the net displacement of the center of topological charge (CTC). This agreement reflects the dominance of isotropic shape deformations, validating the use of $\ell$ and $\mathcal{D}$ as effective shape variables in this regime. For strong coupling, Eq.~\eqref{eq.thiele} fails to capture the full dynamics, indicating that additional soft modes become relevant and couple to the CTC translation, leading to a more complex shape manifold (see End Matter). Fig.~\ref{fig.ResonantSelf-propulsion} (e) also reveals that, for both weak and strong interlayer couplings, the propulsion speed initially scales quadratically with the gyrocoupling-to-friction ratio $\Gamma=G/\alpha\bar{\mathcal{D}}$. However, at high $\Gamma$ (low damping), skyrmions grow too large in one of the half-cycles of the excitation, momentarily favoring the coaxial state, leading to the saturation of $v_\text{sp}$. This saturation can also be seen in Fig.~\ref{fig.ResonantSelf-propulsion} (a) and (c) as smoothened out resonance peaks for large $\Delta K$.

For the ABM case, the situation is considerably more complex. Since $\mathcal{D}_1$ and $\mathcal{D}_2$ oscillate in antiphase, the skyrmion dynamics break mirror symmetry, so $\ell$, $\mathcal{D}_1$, and $\mathcal{D}_2$ must be treated as independent geometric parameters. Moreover, as discussed above, $\ell$ oscillates at twice the excitation frequency due to symmetry constraints, introducing inherent nonlinearity even for small amplitudes and making a simple analytical treatment, as in the SBM case, more challenging. Nonetheless, when projected onto the plane defined by $\ell$ and $\Delta\mathcal{D} \equiv \frac{1}{2}(\mathcal{D}_2 - \mathcal{D}_1)$, the system still traces a closed swimming cycle, as shown in Fig.~\ref{fig.SeriesMechanism} (d). Because $\ell(t)$ oscillates at twice the frequency of $\Delta\mathcal{D}(t)$, the resulting trajectory forms a characteristic 2:1 Lissajous figure~\cite{MarionThornton2003}. Furthermore, the dependence of $v_\text{sp}$ on $\Gamma$ follows the same qualitative trends observed in the SBM case, as shown in Fig.~\ref{fig.ResonantSelf-propulsion} (e).

\begin{figure}[t!]
\includegraphics[width=\linewidth]{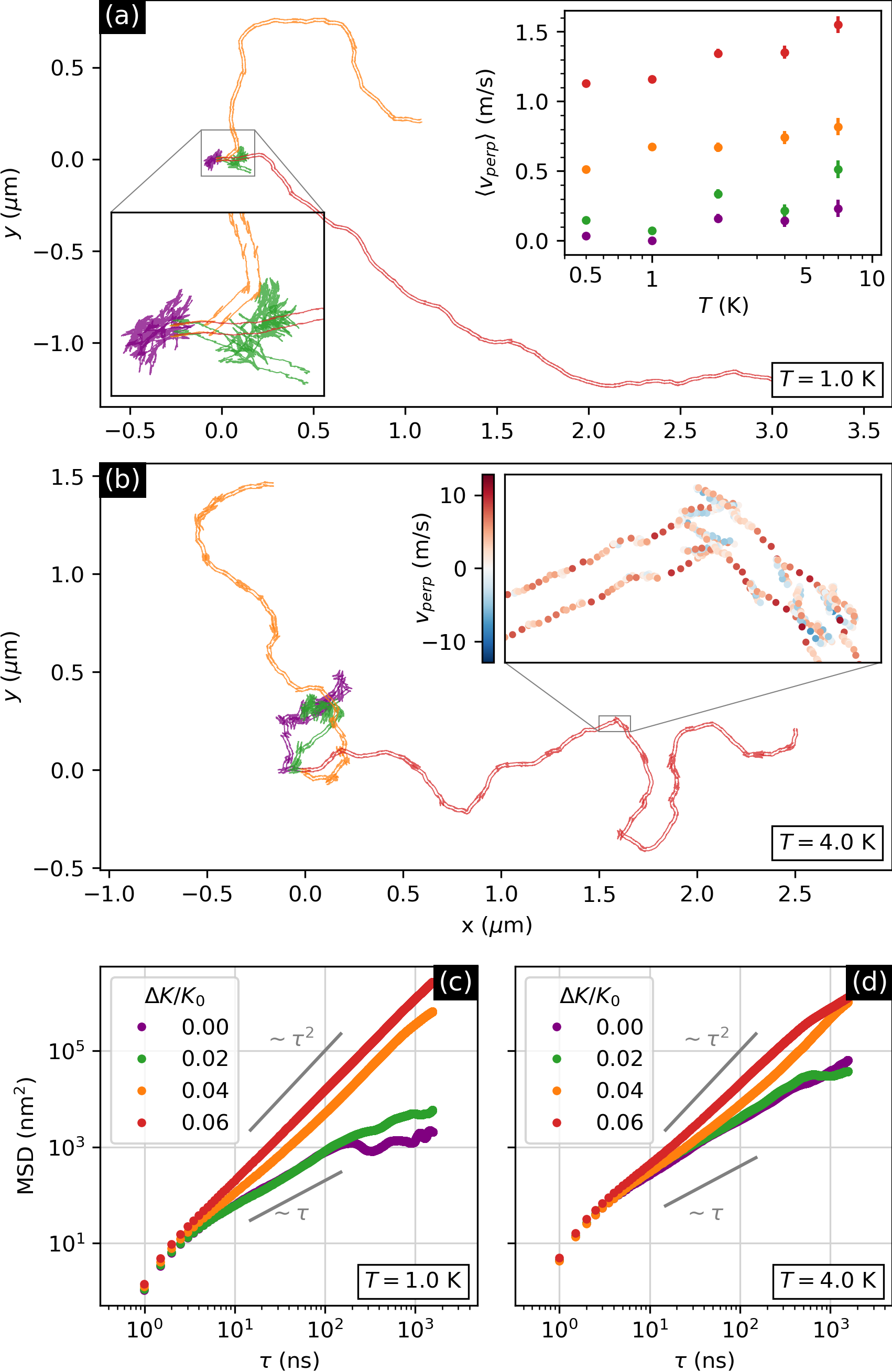}
\caption{(a, b): Typical trajectories of a self-propelled skyrmion pair excited at 20 GHz (near resonance) along $\Delta t=3.116$ $\mu$s, for anisotropy amplitudes $\Delta K/K_0=0$ (purple), 0.02 (green), 0.04 (orange), and 0.06 (red), $K_0=0.35$ MJ/m$^3$, $A_\text{int}=0.12$ mJ/m$^2$, $\alpha=0.1$, and temperatures $T=1.0$ K (a) and 4.0 K (b). In all cases, the pair was initialized at the origin heading west and positions were recorded every 0.5 ns. Zoomed insets in (a) and (b) highlight the low-$\Delta K$ diffusive and high-$\Delta K$ run-and-reverse regimes. In the inset of (b), colors depict the pair velocity perpendicular to the bond, with blueish dots indicating reverse motion. (c, d): Mean-square displacements, ${\rm MSD}(\tau) = \frac{2}{\Delta t}\int_{0}^{\Delta t/2}|\bm{R}(\tau+t)-\bm{R}(t)|^2dt$, corresponding to the trajectories in (a) and (b). The gray lines indicate the diffusive ($\sim \tau$) and ballistic ($\sim \tau^2$) behaviors. Top right inset depicts the average pair velocity component perpendicular to the bond as a function of temperature. 
}
\label{fig.Noise}
\end{figure}

\emph{Influence of thermal noise}---Due to their nanoscale dimensions, skyrmions in ultrathin magnetic materials are highly sensitive to thermal fluctuations, even at temperatures as small as a few Kelvin~\cite{Weissenhoffer2021}. To assess the robustness of the self-propulsion mechanism under realistic conditions, we performed stochastic LLG simulations at temperatures ranging from 0.5 to 7.0 K for strong interlayer coupling. These conditions produce high self-propulsion speeds (up to $\sim1$ m/s) while safely preventing thermal break-up of the pair within the investigated time window. Fig.~\ref{fig.Noise} (a) and (b) shows some representative trajectories of the skyrmion pair subjected to oscillating $K(t)$ and thermal noise~\cite{footnote_Unwrap}.  At low or zero anisotropy amplitudes, the pair jiggles back and forth and changes orientation randomly, producing essentially diffusive dynamics akin to anisotropic Brownian particles~\cite{Han2006,tenHagen2011}. This behavior is confirmed by the linear time dependence of the respective mean-square displacement (MSD) curves shown in panels (c) and (d). In this regime, thermal fluctuations overpower the self-propulsion mechanism, effectively suppressing the net motion. 

At high enough excitation amplitudes ($\Delta K\geq0.04K_0$), the dynamics are markedly different. The skyrmion pair exhibits persistent directed motion over extended distances before reorienting, similar to the active Brownian motion seen in systems such as Janus colloids and motile microorganisms~\cite{Bechinger2016_RMP}. The persistence of motion is reflected in the MSD curves as a quasi-ballistic ($\text{MSD}(\tau)\sim \tau^2$) behavior. While the average propulsion speed seems to slightly increase with temperature, the trajectories become more sinuous at higher $T$, as expected for active Brownian systems. 

A closer examination of the trajectories reveals that, in addition to the slow orientational diffusion, the pair occasionally halts, reverses direction briefly (for a few nanoseconds), and then resumes forward motion, as illustrated in the inset of Fig.~\ref{fig.Noise} (b) (see also the Supplemental Videos~\cite{footnote_SM}). In some cases, the pair performs two or more back-and-forth jitters before restoring persistent forward motion, often emerging along a new direction. Notably, all abrupt changes in orientation coincide with such reversal events.  While these reversals appear to be triggered by fluctuations that transiently push the skyrmions too close to one another, a detailed understanding of the underlying mechanism is still lacking and will be the subject of future investigation. Interestingly, this emergent run-and-reverse pattern is a well documented strategy employed by a large class of motile bacteria to explore complex environments more efficiently~\cite{Blackburn1998,Johansen2002,Xie2011,Son2013,Pablo2020, Anchutkin2024}. In the present context, it could similarly enhance the ability of skyrmion pairs to navigate inhomogeneous or disordered energy landscapes.

\emph{Concluding remarks}---We have shown that bound skyrmion pairs in synthetic antiferromagnets can self-propel at high speeds under ac electric or magnetic fields without the need for external currents or gradients. Related phenomena have been observed in ac-driven bimerons in liquid crystals~\cite{ackerman2017squirming,sohn2019schools}, where motion is guided by the far-field orientation, and in ac-driven single skyrmions, where an additional dc field breaks spatial symmetry~\cite{yuan2019wiggling}. In our case, directed motion emerges instead from local symmetry breaking, through spontaneous, non-reciprocal coupling between shape and translation modes, enabling the skyrmion pair to steer itself. The resulting dynamics are sensitive to thermal fluctuations and show features typical of Brownian active particles, including persistent motion, orientational diffusion, and run-and-reverse behavior, echoing the strategies used by swimming bacteria. 

These findings motivate further exploration of self-propelled skyrmions as components of spin-based active matter, where interactions between autonomous magnetic textures might lead to rich collective behavior. It would also be interesting to investigate how the dynamics evolve in multilayered synthetic antiferromagnets, where the net topological charge depends on the number of layers, and to study interactions among multiple skyrmion pairs. On the application side, the mode-selective nature of the propulsion mechanism could be used to perform programmable tasks using external stimuli, such as ultrashort laser pulses \cite{Titze2024}, that excite specific internal modes. The influence of nanostructured substrates, particularly chiral geometries~\cite{zhang2023chiral}, also offers a promising direction, as it may induce torques or symmetry breaking effects that enrich the propulsion behavior. Combined with feedback control schemes, as recently demonstrated in non-spintronic active matter \cite{fernandez2020}, these skyrmion swimmers may serve as functional elements in low-power neuromorphic computing or stochastic logic devices \cite{Raab2022, song2020skyrmion, Roy2024_Spintronic, liu2022reservoir}.

\bigskip
This work was financed in part by Coordenação de Aperfeiçoamento de Pessoal de Nível Superior – Brasil (CAPES), Finance Code 001, Conselho Nacional de Desenvolvimento Científico e Tecnológico – Brasil (CNPq), Grant No{.} 312240/2021-0, and Fundação de Amparo à Ciência e Tecnologia do Estado de Pernambuco (FACEPE), Grant Number APQ-1129-1.05/24.


\begin{thebibliography}{73}
\bibitem [{{Because $\mathcal{D}$ reflects the breathing mode while $\ell$ is dominated by the binding force, these quantities respond at different rates to the driving field, leading to different phase lags}()}]{footnote_PhaseLag}%
  \BibitemOpen
  {Because $\mathcal{D}$ reflects the breathing mode while $\ell$ is dominated by the binding force, these quantities respond at different rates to the driving field, leading to different phase lags},\ \href@noop {} {}\BibitemShut {NoStop}%
\bibitem [{{To clearly show the pair's movement as if it were in an infinite plane, we ``unwrap" its path. This means we add or subtract the simulation box size (100 nm) whenever it crosses a boundary, creating a continuous trajectory}()}]{footnote_Unwrap}%
  \BibitemOpen
  {To clearly show the pair's movement as if it were in an infinite plane, we ``unwrap" its path. This means we add or subtract the simulation box size (100 nm) whenever it crosses a boundary, creating a continuous trajectory},\ \href@noop {} {}\BibitemShut {NoStop}%
\end{thebibliography}

%

\newpage

\onecolumngrid  

\section{End Matter}

\twocolumngrid
\setcounter{equation}{0}
\appendix

\renewcommand{\theequation}{A\arabic{equation}}

\hypertarget{app.Numerics}{\emph{Appendix A: Numerical details}}---We model the dynamics of the local normalized magnetization of each layer, $\bm{m}_i(t)=\bm{M}_i(t)/M_s$, by the stochastic LLG equation:
\begin{equation}
\begin{aligned}
\label{eq.stochasticLLG}
\frac{\partial\bm{m}_i}{\partial t} =&-\frac{\gamma}{1+\alpha^2}\{\bm{m}_i\times\left(\bm{B}_{\text{eff},i}+ \bm{B}_\text{th}\right)\\ &+\alpha\bm{m}_i\times\left[\bm{m}_i\times\left(\bm{B}_{\text{eff},i}+ \bm{B}_\text{th}\right)\right]\}\textrm{.}
\end{aligned}
\end{equation}
The stochastic field $\bm{B}^{\text{th}}(\bm{r},t)$ models thermal fluctuations at temperature $T$ and satisfies $\langle {B}^{\text{th}}_{j}(\bm{r},t)\rangle = 0$ and 
\begin{equation}
\begin{aligned}
    \left \langle {B}^{\text{th}}_{k}(\bm{r},t){B}^{\text{th}}_{l}(\bm{r}',t') \right \rangle = 2\lambda_{kl}\delta(\bm{r}-\bm{r}')\delta(t-t') \text{,}
\end{aligned}
\end{equation}
where $k$, $l$ are the cartesian components, $\lambda_{kl}=(\alpha k_BT/\gamma M_s)\delta_{k,l}$, $k_B$ is the Boltzmann's constant, and $\langle\,\rangle$ represents the ensemble average~\cite{NagaosaStochastic2014}. The effective field $\bm{B}_{\text{eff},i}(t)= -M_s^{-1} (\delta\mathcal{H}/\delta\bm{m}_i)$ is calculated via the functional derivative of the total micromagnetic Hamiltonian
\begin{equation}
\label{eq.Hamiltonian}
    \mathcal{H} = \int dS\left[\mathcal{E}_1 d + \mathcal{E}_2 d + A_\text{int}\bm{m}_1\cdot\bm{m}_2 \right]\textrm{,}
\end{equation}
where $A_\text{int}>0$ is the antiferromagnetic interlayer coupling stiffness and $\mathcal{E}_i$ is the intralayer energy density of the $i$th layer. $\mathcal{E}_i$ includes exchange, anisotropy, Zeeman, and Dzyaloshinskii–Moriya (DM) contributions as %
\begin{align} 
\label{eq.IntralayerInteractions}    \mathcal{E}_i &= A\left[\left(\partial_x\bm{m}_i\right)^2 + \left(\partial_y\bm{m}_i\right)^2\right] - K m_{z(i)}^2 -M_s \bm{B}\cdot \bm{m}_i  \nonumber \\
    & + D_{i}\left[m_{\mathrm{z}(i)}\left(\bm{\nabla}\cdot\bm{m}_i\right)-\bm{m}_i\cdot\left(\bm{\nabla} m_{\mathrm{z}(i)}\right)\right]\textrm{.} 
\end{align}
The exchange stiffness of both layers is $A=15$ pJ/m and their DM constants $D_1=-D_2=3.05\,\text{mJ/m²}$. $K=K_0+\Delta K$ is the effective anisotropy parameter, where $K_0=K_u+\frac{1}{2}\mu_0M_s^2$ accounts for the non-modulated uniaxial perpendicular anisotropy and demagnetization effects~\cite{EffectiveAnisotropy_ThinFilms}, and $\Delta K$ is the (linear response) change in perpendicular anisotropy due to a perpendicular electric field~\cite{upadhyaya2015electric}. 
\begin{figure}[t!]
\includegraphics[width=\linewidth]{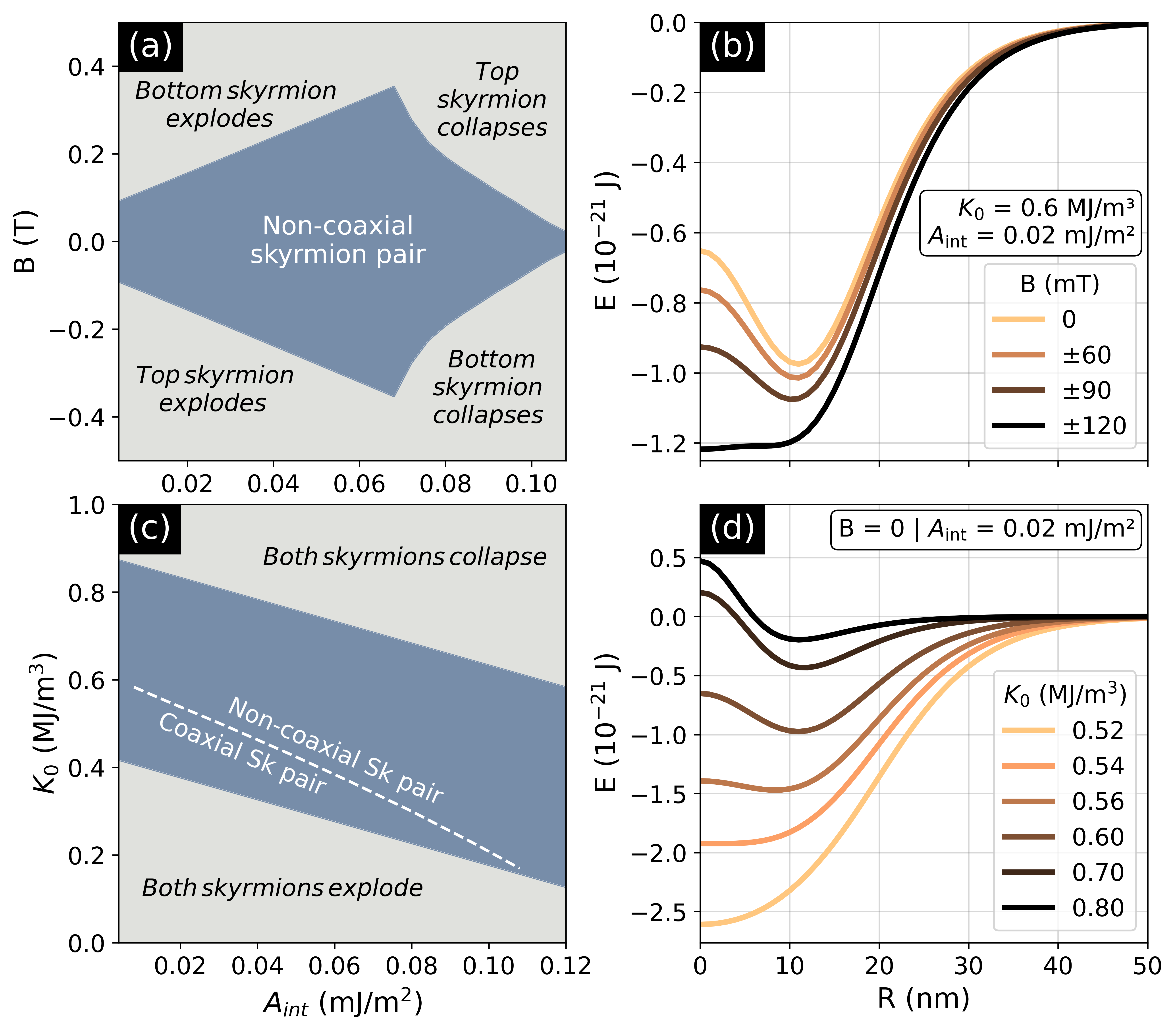}
\caption{Phase diagrams and skyrmion-skyrmion potential profiles obtained by varying  $A_\text{int}$ and either the anisotropy $K_0$ [(a)-(b)] or the dc magnetic field $\bm{B} = B\hat{\bm{z}}$ [(c)-(d)], whilst fixing the other parameters. (a) For $K_0=0.6$ MJ/m³, the non-coaxial skyrmion pair is stable for a symmetric range of $B$ with respect to $B=0$ (dark region) and unstable outside this region, with one of the skyrmions collapsing or exploding. (c) For $B=0$ and varying $K_0$, the stability region of the pair now features both possibilities, coaxial and non-coaxial, separated by the dashed line. Panels (b) and (d) show selected energy profiles, $E(R)$, for $A_\text{int}
= 0.02\, \text{mJ/m²}$. The equilibrium bond length of the pair can be identified as the position $R$ of minimum $E$.}
\label{fig.EquilibriumPhaseDiagram}
\end{figure}
We solve Eq.~\eqref{eq.stochasticLLG} on a $100\times100$ nm$^2$ grid per layer, discretized into cells of size $1\times1$ nm$^2$, with periodic boundary conditions using a second order stochastic Runge-Kutta method. 

To find the possible equilibrium configurations of the skyrmion pair, we initialize the FM layers with opposite magnetization background, and introduce a prototype skyrmion in each layer using a 360$^\circ$ domain-wall ansatz~\cite{TheorySKSize}. The resulting skyrmions have opposite charges and chiralities and are initially separated by a distance of 10 nm. The system is then relaxed at $T=0$, allowing the skyrmions to reach their natural shape and equilibrium separation or, in some cases, to undergo collapse or explosion. The resulting configurations are summarized in Fig.~\ref{fig.EquilibriumPhaseDiagram}~(a) and (c), as a function of $A_\text{int}$, $\mathbf{B}=B \hat{\mathbf{z}}$ and $K_0$. Selected energy profiles, computed following the procedure described in~\cite{Matheus2024}, are shown in panels (b) and (d), illustrating how the bond length evolves with the control parameters.  \\

\setcounter{equation}{0}

\bigskip
\renewcommand{\theequation}{B\arabic{equation}}

\begin{figure}[t!]
\includegraphics[width=\linewidth]{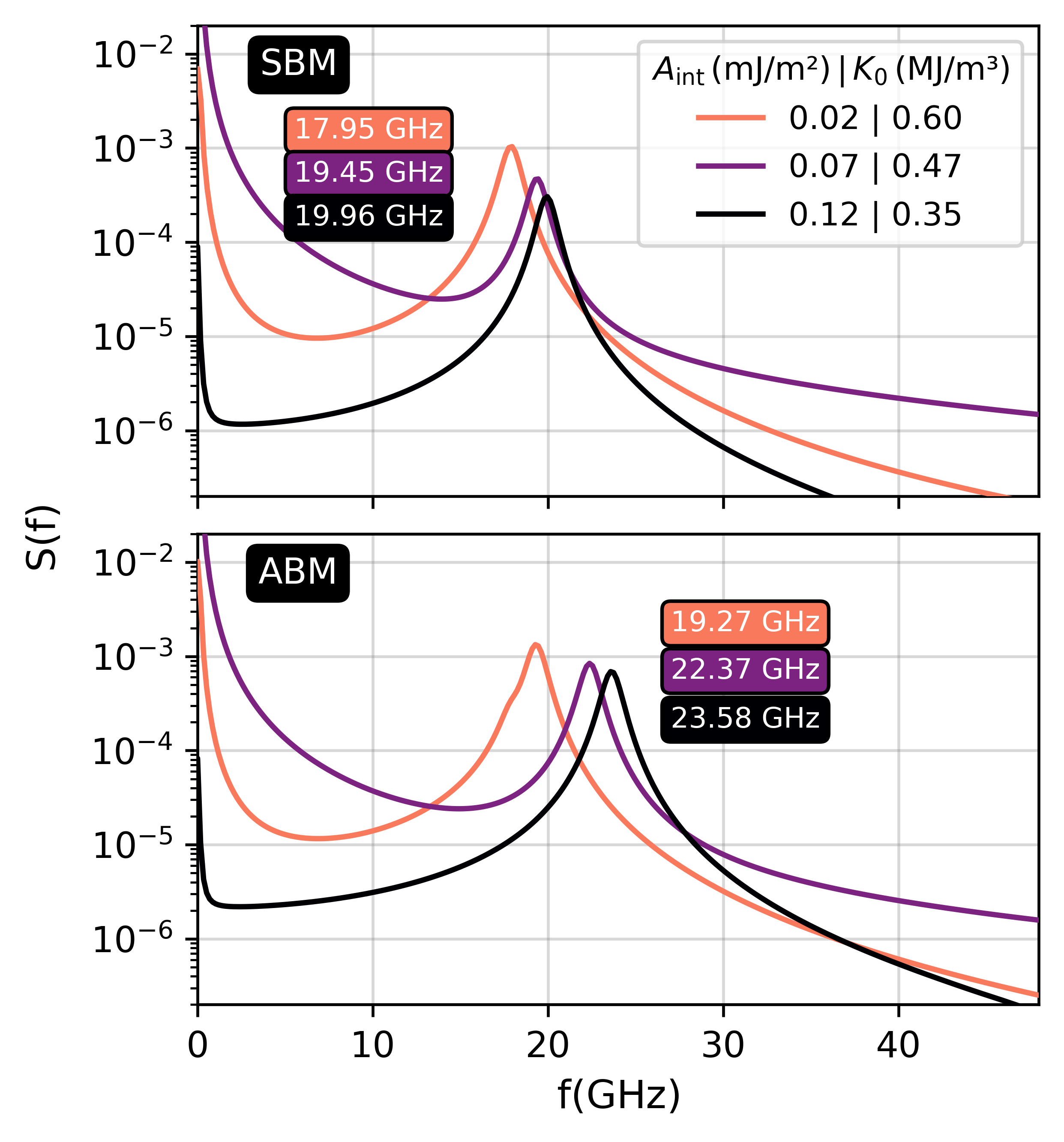}
\caption{Power spectra of the magnetization response to time varying anisotropy (SBM) and magnetic field (ABM) computed for damping $\alpha=0.02$. Weak, intermediate, and strong coupling regimes are represented, respectively, by $A_\text{int}=0.02$, 0.07, and $0.12$ mJ/m², with a corresponding decrease in $K_0$ to ensure the pair remains well within the non-coaxial phase shown in Fig.~\ref{fig.EquilibriumPhaseDiagram}-(c). For SBM excitation, the anisotropy is modulated with amplitude $\Delta K/K_0 = 0.004$ at zero magnetic field. For ABM excitation, the magnetic field is applied with amplitude $B_0 = 4$ mT while keeping $K = K_0$ fixed. Resonance frequencies $f_R$ are indicated next to each curve.
}
\label{fig.SpectraABM_SBM}
\end{figure}

\hypertarget{app.NormalModes}{\emph{Appendix B: Spectral analysis}}---We analyze the breathing modes following a procedure similar to that in Ref.~\cite{SkBreathingModes}. At $T=0$, we compute the spatially averaged magnetization $\left\langle m_z\right\rangle = (1/V)
\int m_z(\bm{r})\,dV$ for each FM layer at equilibrium, denoted $\left\langle m_z\right\rangle_\text{eq}$. To excite the system we apply either a time-varying magnetic field $\bm{B}_{\text{ext}}(t) = \mathbf{\hat{z}}B_0 \text{sinc}{(f_c t)}$ or anisotropy parameter $K(t) = K_0+ \Delta K \text{sinc}{(f_c t)}$, with amplitudes $B_0=4$ mT and $\Delta K=0.004K_0$, and cutoff frequency $f_c=100\,\text{GHz}$. The cardinal sine shape of the excitations, $\text{sinc}(x)=\sin(\pi x)/\pi x$, ensures uniform excitation across frequencies up to $f_c$. We record the deviation of the mean magnetization from equilibrium, $\delta m_z(t) = \left\langle m_{z}(t)\right\rangle -  \left\langle m_z\right\rangle_\text{eq}$, every 2 ps throughout $\Delta t=10$ ns. The excited modes are identified from the power spectrum
$S(f) = \left| \int_{0}^{\Delta t} dt \, e^{-i 2\pi f t} \delta m_z(t) \right|^2$. 
Figure~\ref{fig.SpectraABM_SBM} 
shows spectra for several values of $A_\text{int}$ and $K_0$, covering weak to strong coupling. The top panel corresponds to $K(t)$ excitations, which selectively excite symmetric breathing modes (SBM), while the bottom panel corresponds to $\bm{B}_{\text{ext}}(t)$ excitations, which couple to antisymmetric breathing modes (ABM). The position of the resonance peaks, $f_R$, is observed to increase with $A_\text{int}$, and the ABM spectra exhibit slightly higher $f_R$ values.
\bigskip
\renewcommand{\theequation}{C\arabic{equation}}

\hypertarget{app.Thiele}{\emph{Appendix C: Generalized Thiele equation}}---In the presence of time-varying fields or inhomogeneities, soft shape modes of the skyrmion can be excited, including breathing and anisotropic deformations. To incorporate these effects in a Thiele-like approach, the LLG dynamics can be projected onto a space spanned by both translation and a suitable set of shape modes, $\{\xi_1,\xi_2,\dots\}$, resulting in generalized Thiele equations~\cite{Tretiakov2008,Liu2023}. For translation, this equation reads
\begin{equation}\label{eq.GenThiele}
    \bm{G}\times\dot{\bm{R}} + \alpha\bm{\mathcal{D}}\dot{\bm{R}} = \bm{F} + \sum_j \begin{pmatrix} \mathcal{G}_{xj}-\alpha \mathcal{D}_{xj} \\ \mathcal{G}_{yj}-\alpha \mathcal{D}_{yj} \end{pmatrix}\dot{\xi}_j,
\end{equation}
where $\mathcal{G}_{\mu j}=\frac{dM_s}{\gamma}\int d^2r\,\bm{m}\cdot\left(\frac{\partial\bm{m}}{\partial \mu} \times \frac{\partial\bm{m}}{\partial \xi_j}\right)$ and $\mathcal{D}_{\mu j}=\frac{dM_s}{\gamma}\int d^2r\,\frac{\partial\bm{m}}{\partial \mu} \cdot \frac{\partial\bm{m}}{\partial \xi_j}$ are respectively the gyrotropic and dissipative coupling coefficients between the translation mode $\mu$ ($=x,y$) and the shape mode $\xi_j$. Notably, isotropic modes such as the skyrmion radius $\mathcal{R}$ do not contribute explicitly to the right-hand side of Eq.~\ref{eq.GenThiele}, because, by symmetry, $\mathcal{G}_{\mu\mathcal{R}}$ and $\mathcal{D}_{\mu\mathcal{R}}$ vanish. However, the dynamics of $\mathcal{R}(t)$ directly modulates the dissipation tensor $\bm{\mathcal{D}}$. In this special case, Eq.~\ref{eq.GenThiele} reduces to Eq.~\ref{eq.thiele} with time-dependent $\bm{\mathcal{D}}$.

Applying Eq.~\ref{eq.GenThiele} to the SBM dynamics of a SAF skyrmion pair and assuming perfect mirror symmetry, we get
\begin{equation}\label{eq.vperp_general}
    v_\perp = -\frac{G-\alpha\mathcal{D}_{xy}}{2\alpha\mathcal{D}_{yy}}\dot{\ell} + \sum_j\frac{\mathcal{G}_{yj}-\alpha\mathcal{D}_{yj}}{\alpha\mathcal{D}_{yy}}\dot{\xi}_j,
\end{equation}
where the $x$ and $y$ axes are defined respectively parallel and perpendicular to the skyrmion bond length $\ell$. This equation allows to express the self-propulsion speed as a closed-path integral in the extended shape space $\{\ell,\xi_1,\xi_2,\dots\}$. For circular skyrmions ($\mathcal{D}_{xx}=\mathcal{D}_{yy}=\mathcal{D}$ and $\mathcal{D}_{xy}=0$) undergoing pure (isotropic) breathing, the second term in Eq.~\ref{eq.vperp_general} vanishes and $v_\text{sp}$ reduces to the simple form given in Eq.~\eqref{eq.vsp}. In contrast, when an external field excites other shape modes that couple non-reciprocally to translation, Eq.~\eqref{eq.vsp} becomes insufficient. In such cases, the full dynamics of Eq.~\eqref{eq.vperp_general} must be considered to properly account for shape–translation couplings contributing to propulsion.

\newpage

\onecolumngrid  

\section*{Supplemental Material}

\setcounter{equation}{0}
\renewcommand{\theequation}{S\arabic{equation}}

\section{Description of the supplemental videos}
\label{sec.Videos}

To better visualize the self-propulsion dynamics at long time scales, we provide several Supplemental Videos showing the time evolution of the spin configuration, obtained from the micromagnetic simulations, and the resulting trajectories of the skyrmion pair at different temperatures. In all cases, the skyrmion pair is initially placed at the lower center of a 100$\times$100 nm$^2$ simulation box with periodic boundary conditions. The initial bond vector $\bm{\ell}=\bm{R}_1-\bm{R}_2$ points to the left, so that the initial heading of the pair points upwards. The pair is then excited in the symmetric breathing mode (SBM) under the same conditions as those used in Fig. 4 of the main text. Each video spans a physical time between 1.3 and 2 $\mu$s, with a frame interval of 5 ns. The final frames of each video are shown in Fig.~\ref{fig.SuppVideos} for reference. Further details of each video are given below.

\begin{itemize}

    \item {\bf SuppVideo1:} Time evolution of the spin configuration, represented by the total out-of-plane magnetization $m' = m_{z,1} + m_{z,2}$ at zero temperature ($T = 0$ K), for different excitation amplitudes $\Delta K/K_0$ at 20 GHz. The parameters are $A_{\text{int}} = 0.12$ mJ/m², $K_0 = 0.35$ MJ/m³, and $\alpha = 0.1$. Red (blue) indicates regions with $m' = +1$ ($-1$), while gray corresponds to $m' = 0$. (e--h): Corresponding trajectories of the skyrmion pair for $\Delta K/K_0 = 0$ (purple), 0.02 (green), 0.04 (yellow), and 0.06 (red). Unlike magnetization maps, which are confined to the simulation box, the trajectories are displayed with periodic boundary conditions unwrapped, allowing the pair's path to be visualized continuously across the sample. For nonzero excitation, the pair exhibits straight-line motion perpendicular to the bond, with the self-propulsion speed $v_\text{sp}$ increasing monotonically with $\Delta K/K_0$.

    \item {\bf SuppVideo2:} Same as SuppVideo1, but at a finite temperature $T=1$ K. For $\Delta K/K_0 = 0$ and 0.02, thermal fluctuations dominate, and the pair diffuses within a region of about  0.2$\times$0.2 $\mu$m$^2$ near the origin. At higher excitations ($\Delta K/K_0 = 0.04$ or 0.06), self-propulsion becomes strong enough to overcome thermal diffusion, and the pair enters a ballistic regime, covering much longer distances. However, thermal noise still introduces deviations from a perfectly straight path. 
    
    \item {\bf SuppVideo3:} Same as SuppVideo1, but at $T=4$K. For weak or no excitation, the pair remains near the origin, with enhanced diffusion due to the higher temperature. For $\Delta K/K_0 = 0.04$ and 0.06, we observe ballistic motions intercalated with spontaneous reversals of the propulsion speed, resulting in a run-and-reverse pattern akin to behaviors observed in some microswimmers. 
\end{itemize}

\begin{figure}[p]
\includegraphics[width=0.75\linewidth]{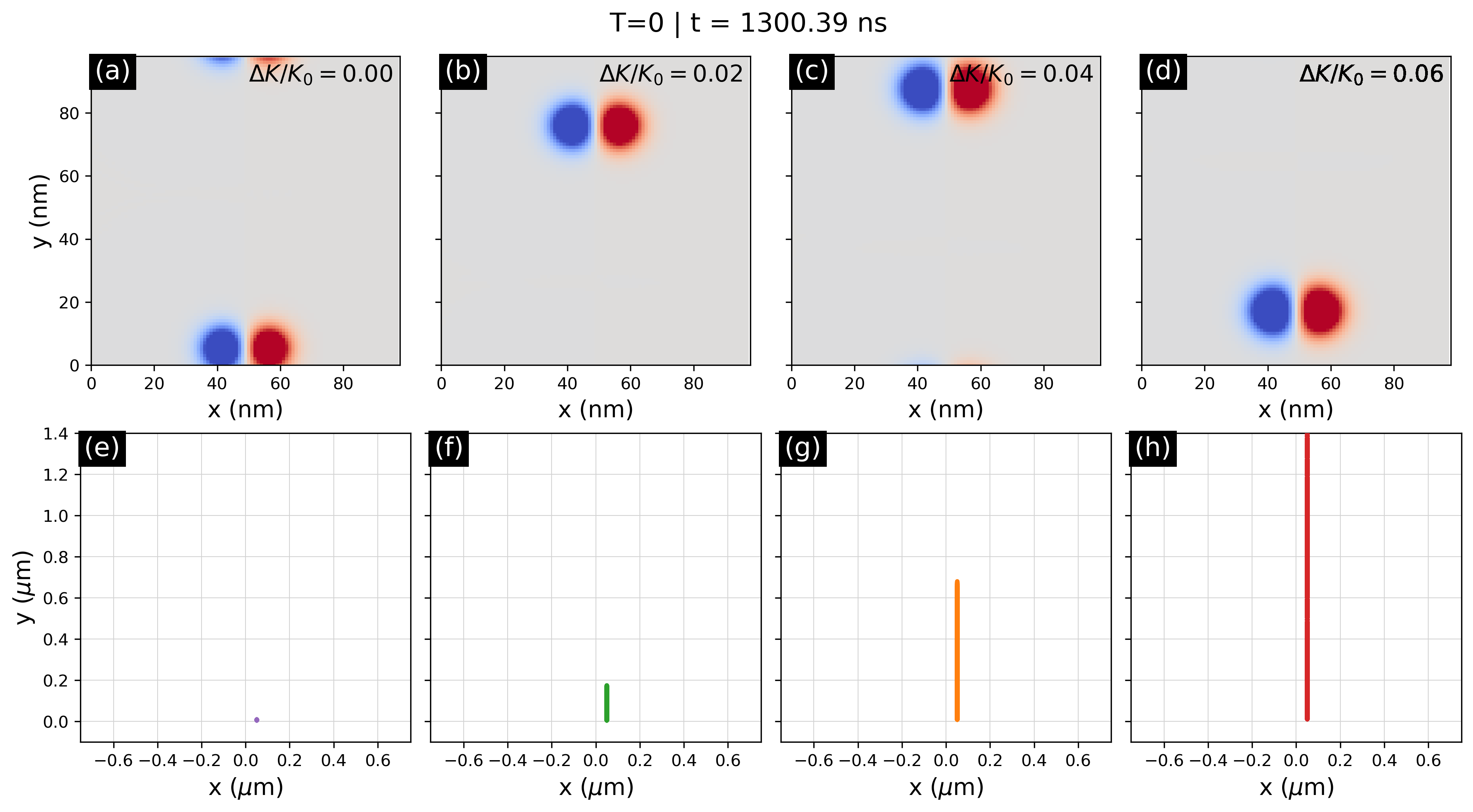}
\includegraphics[width=0.75\linewidth]{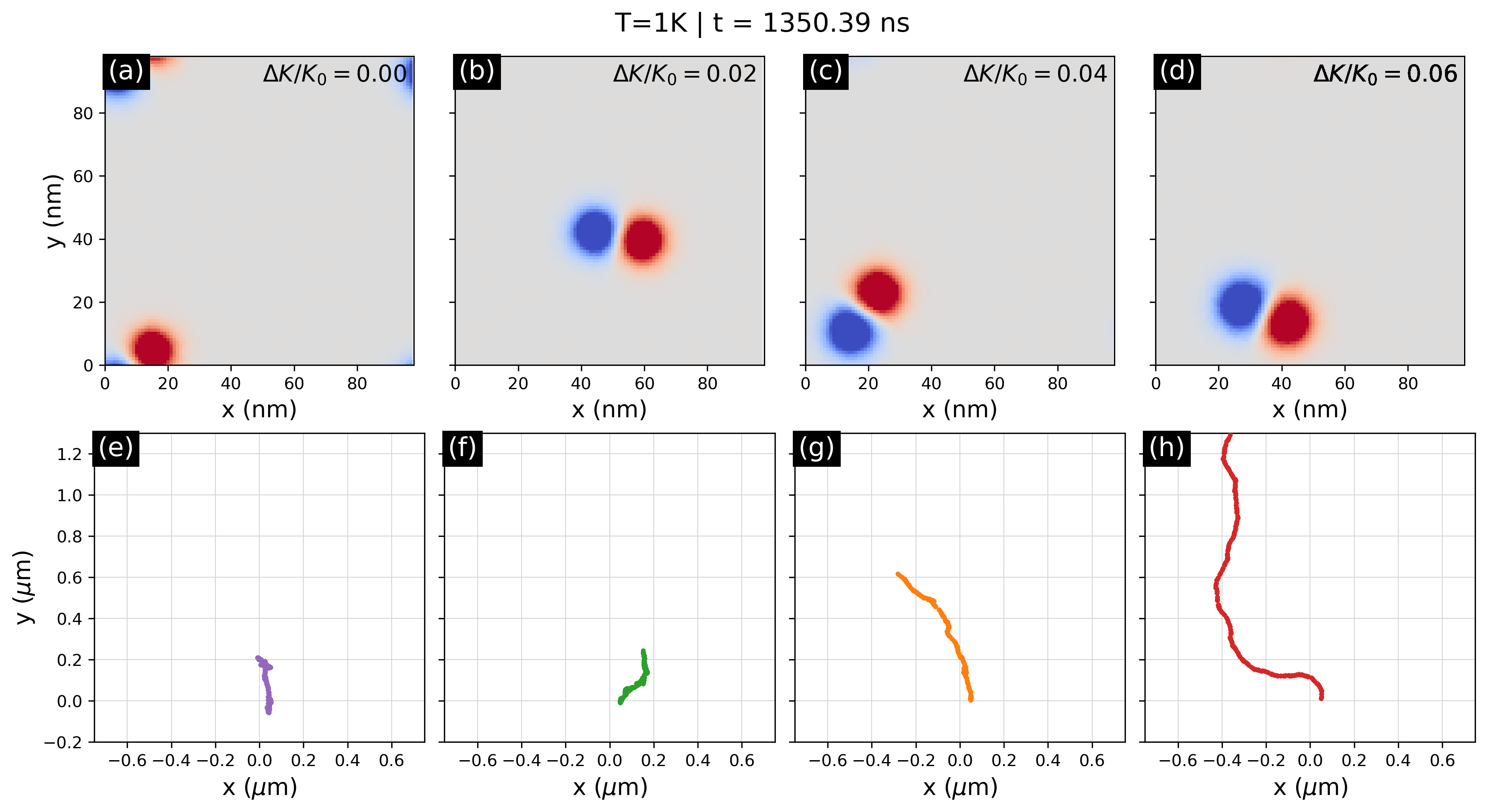}
\includegraphics[width=0.75\linewidth]{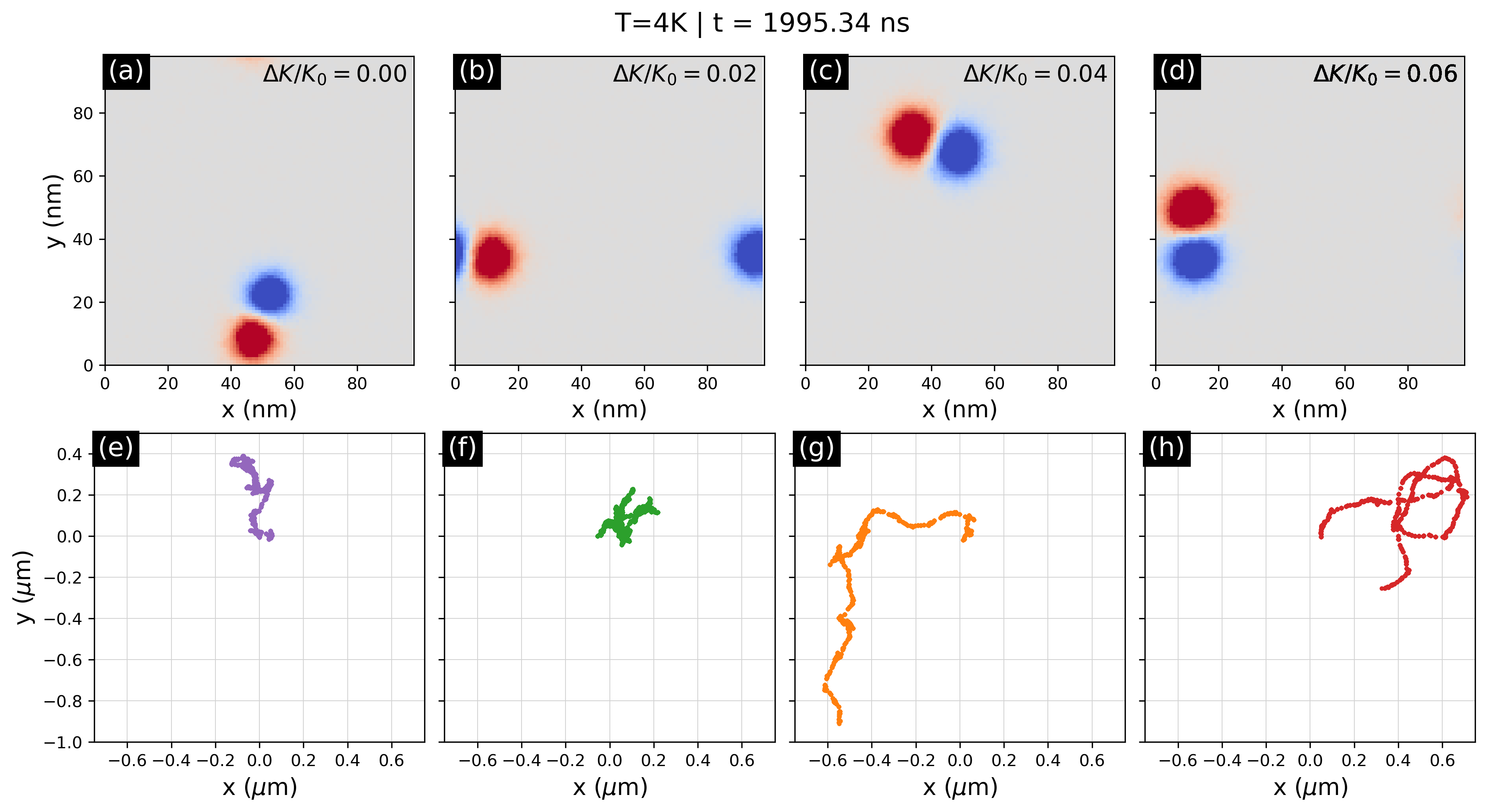}

\caption{Last frames of the supplemental videos.}

\label{fig.SuppVideos}
\end{figure}

\section{Interlayer coupling and skyrmion-skyrmion pair potential}
\label{sec.InterlayerCaoupling}

The stability of a non-coaxial skyrmion pair in a synthetic antiferromagnet (SAF) depends in a nontrivial way on several parameters. Varying only one parameter, such as interlayer coupling strength, while keeping the others fixed, might lead to undesirable results. For example, as suggested by Fig. 5(c), increasing $A_\text{int}$ at constant $K_0$ can destabilize the pair and even cause skyrmion collapse. In contrast, Fig. 5(c) indicates that maintaining stability as $A_\text{int}$ increases requires a simultaneous reduction of the anisotropy constant $K_0$, so as to keep the pair well within the stable region. 

\begin{figure}[t]
\includegraphics[width=0.5\linewidth]{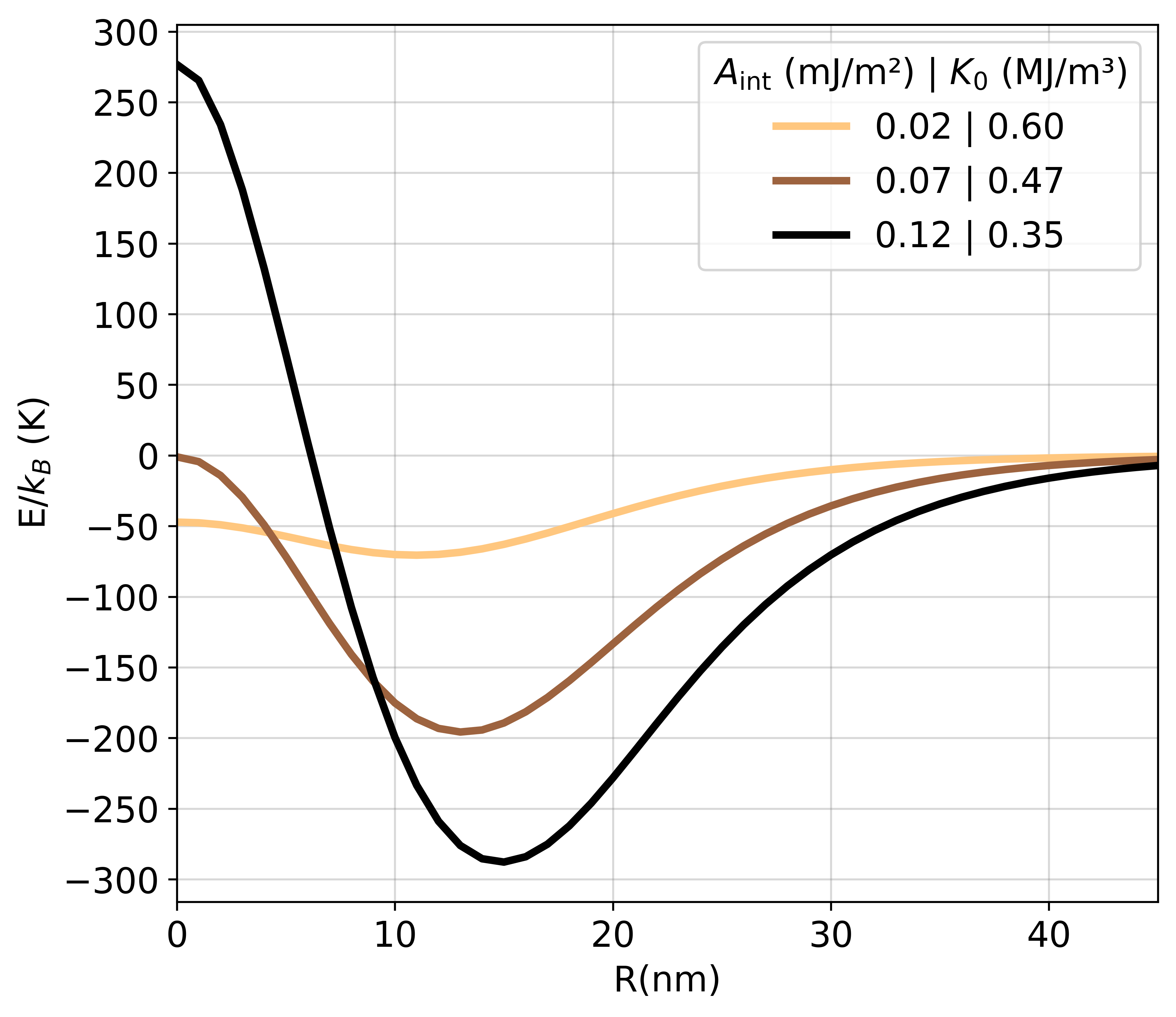}
\caption{Skyrmion-skyrmion interaction potential as a function of distance calculated using the same procedure as in Fig. 5 for three different coupling regimes. The values of $A_\text{int}$ and $K_0$ were chosen according to Eq.~\eqref{eq.KoAint}.
}
\label{fig.Profiles}
\end{figure}

Here, we investigate the evolution of the skyrmion-skyrmion interaction potential as $A_\text{int}$ is varied while $K_0$ is adjusted according to the empirical relation
\begin{equation} \label{eq.KoAint}
  K_0=0.65-2.5A_\text{int},  
\end{equation}
with $K_0$ in MJ/m$^3$ and $A_\text{int}$ in mJ/m$^2$. This equation approximately traces the centerline of the stability region. Skyrmion-skyrmion energy profiles for three points along this line are shown in Fig.~\ref{fig.Profiles}, demonstrating that the interaction becomes considerably stronger as one increases $A_\text{int}$ and decreases $K_0$ according to Eq.~\eqref{eq.KoAint}. To gain further intuition about the effect of thermal fluctuations on pair stability, we express the interaction energies in units of Kelvin. For the weak coupling case, the energy barriers for coaxial alignment and pair dissociation are $\sim20$ K and $\sim70$ K, respectively, rendering the pair highly sensitive to thermal noise even at a few Kelvin. In contrast, for stronger coupling, the energy barriers become substantially larger, resulting in much greater thermal resilience of the skyrmion pair.

\section{Dissipation tensor beyond the weak coupling limit}
\label{sec.DissipationTensor}

Under appropriate conditions, an isolated skyrmion can exhibit a pure isotropic breathing mode. In a skyrmion pair, however, mutual interaction generally breaks the radial symmetry, inducing anisotropic deformations that can be captured in part by the components of the dissipation tensor. As discussed in the main text, the isotropy condition ($\mathcal{D}_{xx}=\mathcal{D}_{yy}=\mathcal{D}$ and $\mathcal{D}_{xy}=0$) is a crucial step for deriving the $\ell$-$\mathcal{D}$ shape integral form of the propulsion speed, Eq. (4). To assess deviations from isotropy, we calculated the steady state time evolution of all three components of the dissipation tensor, $[\bm{\mathcal{D}}_i]_{\mu\nu} = \frac{M_sd}{\gamma}\int d^2r\,\partial_\mu\bm{m}_i \cdot \partial_\nu\bm{m}_i$, under different coupling conditions. For that, we varied $A_\text{int}$ and $K_0$ following Eq.~\eqref{eq.KoAint}.

\begin{figure}[t]
\includegraphics[width=\linewidth]{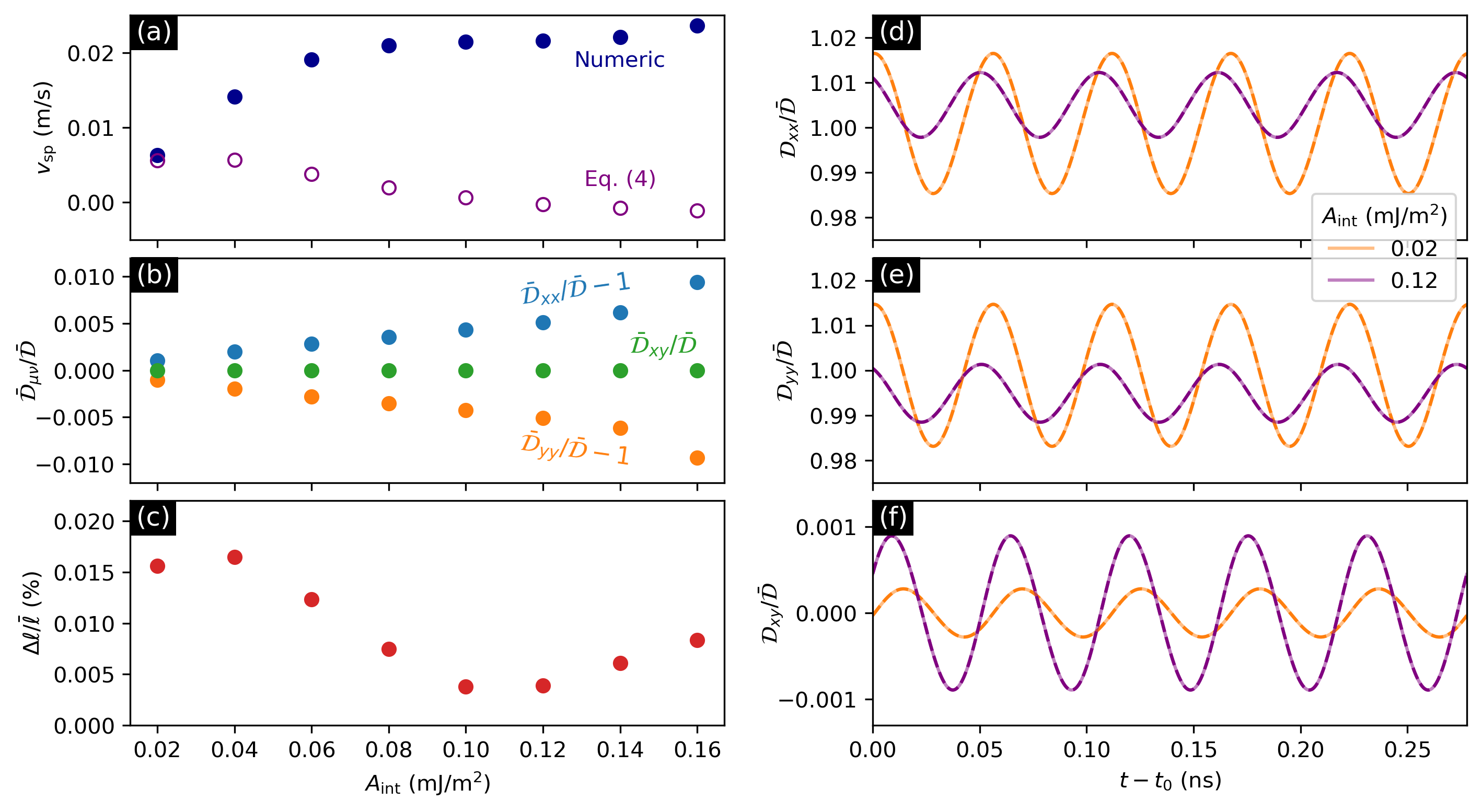}
\caption{(a) Self-propulsion speed, $v_\text{sp}$, (b) normalized time-averaged dissipation tensor components of the top skyrmion, $\bar{\mathcal{D}}_{\mu\nu}/\bar{\mathcal{D}}$, relative to their isotropic limit values, and (c) relative amplitude of the bond length, $\Delta\ell/\bar{\ell}$, as a function of interlayer coupling $A_\text{int}$. In (a), closed symbols represent the propulsion speed calculated directly from the simulated skyrmion displacements, while open dots correspond to the weak-coupling estimate, Eq. (4). (d)--(f) Steady-state time evolution of the dissipation tensor components of the top (semi-transparent solid lines) and bottom (dashes) skyrmion, $\mathcal{D}_{xx}$ (d), $\mathcal{D}_{yy}$ (e), and $\mathcal{D}_{xy}$ (f), for weak ($A_\text{int}=0.02$ mJ/m$^2$, $K_0=0.6$ MJ/m$^3$) and strong ($A_\text{int}=0.12$ mJ/m$^2$, $K_0=0.35$ MJ/m$^3$) couplings. $t_0=38.95$ ns.
}
\label{fig.Dxy}
\end{figure}

Fig.~\ref{fig.Dxy}(a)--(c) shows the dependence of the propulsion speed, time-averaged dissipation tensor components, and bond-length deformation on $A_\text{int}$ [with $K_0$ given by Eq.~\eqref{eq.KoAint}] for SBM excitation at 18 GHz with $\Delta K=0.01K_0$ and $\alpha=0.1$.
Under weak interlayer coupling, the violation of the isotropic condition is small, so the simple isotropic model of the main text provides a satisfactory explanation of the swimming mechanism. For stronger coupling, the isotropic model fails to capture the correct propulsion speed, indicating that anisotropic deformations of the skyrmions might play a major role in the propulsion mechanism. Indeed, in panel (b), one observes the systematic increase of $\bar{\mathcal{D}}_{xx}$ and decrease of $\bar{\mathcal{D}}_{yy}$, with $(\bar{\mathcal{D}}_{xx}-\bar{\mathcal{D}}_{yy})/\bar{\mathcal{D}}$ varying from $\sim0.1\%$ to $\sim1\%$ as $A_\text{int}$ increases from 0.02 to 0.16 mJ/m$^2$, while $\bar{\mathcal{D}}_{xy}$ remains essentially zero. This points to an average elongation of the skyrmions in the ($x$) direction perpendicular to the bond as interlayer coupling increases. At the same time, the bond length deformation presents a nonmonotonic dependence on $A_\text{int}$. While this might suggest that skyrmion elongation somehow dominates the propulsion mechanism, a closer look at the time dependence of the dissipation tensor components, Fig.~\ref{fig.Dxy}(d)--(f), reveals a more intricate dynamics. First, notice that the time-dependent dissipation tensors of the top (continuous line) and bottom (dashed line) skyrmions are identical in both weak and strong coupling regimes. This reveals that the mirror symmetry is robust against the increase of $A_\text{int}$. Now, looking at the tensor components comparatively, one observes that $\mathcal{D}_{xy}(t)$ oscillates with an amplitude that generally increases with $A_\text{int}$, while the amplitudes of $\mathcal{D}_{xx}(t)$ and $\mathcal{D}_{yy}(t)$ decrease. An oscillation of $\mathcal{D}_{xy}$ can be interpreted as the main anisotropy axis wobbling around the direction perpendicular to the bond. Whether this wobbling can couple to the pair translation and account for the propulsion speed beyond the weak coupling regime is still unclear and will be the subject of further investigation.

\begin{figure}[t]
\includegraphics[width=0.6\linewidth]{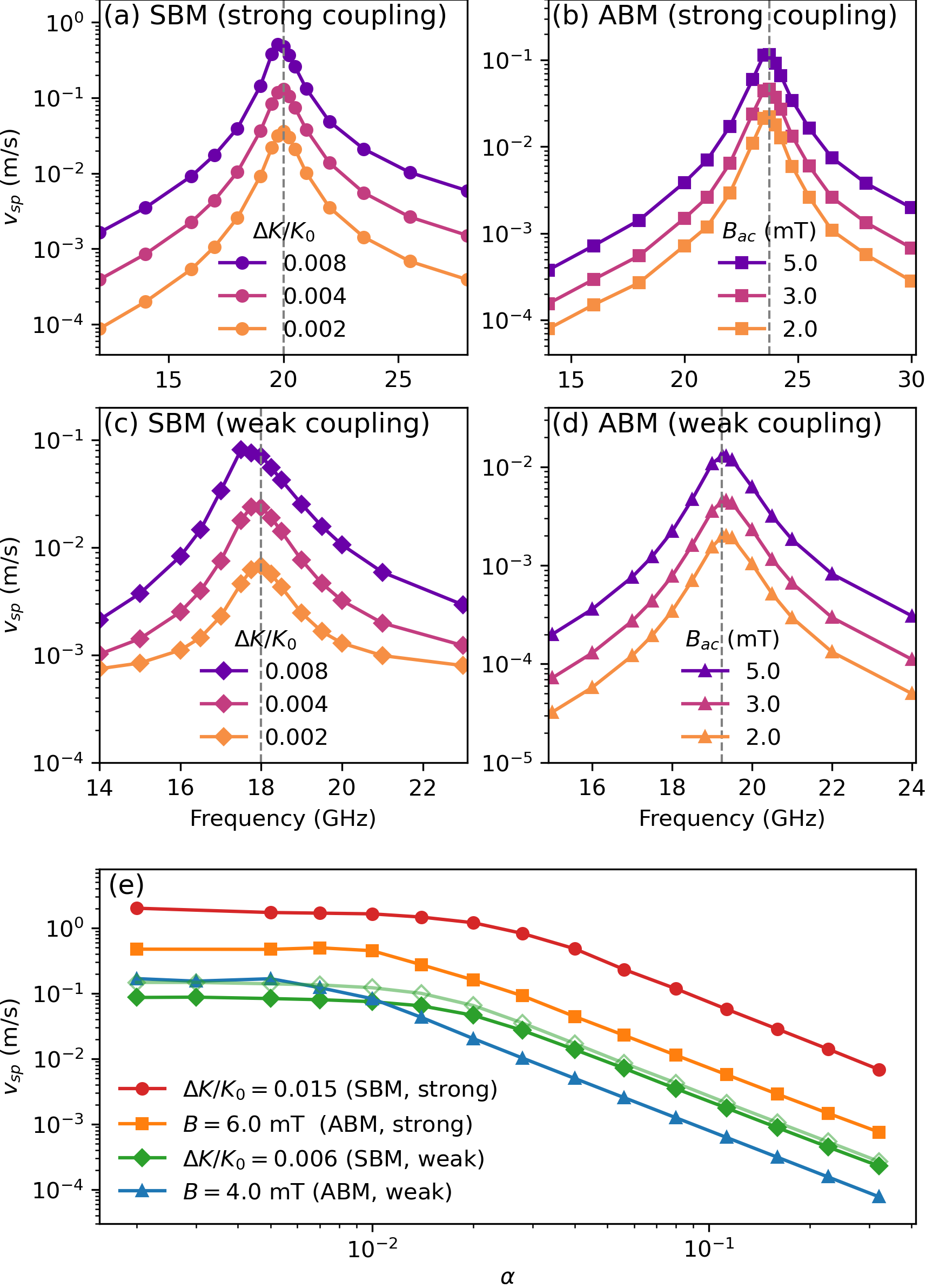}
\caption{(a)-(d) Same as Fig.~2(a)--(d) of the main text, but for the simulations performed under small damping ($\alpha=0.02$). (e) Same as Fig.~2(e), but presenting the data directly as a function of $\alpha$.
}
\label{fig.vspSM}
\end{figure}

\section{Effect of Gilbert damping }
\label{sec.GilbertDamping}

Magnetic multilayers with interfacial Dzyaloshinskii–Moriya interaction between the FM and HM layers typically exhibit Gilbert damping in the range $0.01<\alpha<0.3$~\cite{Liu2024,flacke2021}. 
This allows for a wide span of gyroforce-to-friction ratios ($\Gamma=G/\alpha\bar{\mathcal{D}}$), enabling a broad tunability of the propulsion speed for a given excitation amplitude, as shown in Fig.~2(e). For clarity, we reproduce that figure in Fig.~\ref{fig.vspSM}(e), this time plotting $v_\text{sp}$ directly as a function of $\alpha$, to better visualize how the propulsion speed varies with damping.  

To further explore the impact of the damping on the resonant behavior of $v_\text{sp}$, we show in Fig.~\ref{fig.vspSM}(a)--(d) results analogous to those of Fig.~2(a)--(d) of the main text, but obtained using $\alpha=0.02$, instead of the larger value $\alpha=0.1$ used in the main text. As expected, the resonance peaks are considerably narrower at low damping. Furthermore, since lower damping enhances the efficiency of the breathing-mode excitation, smaller excitation amplitudes are sufficient to achieve a given propulsion speed. In contrast, the maximum propulsion speed is approximately the same for both damping regimes. This saturation arises because, beyond a threshold in breathing amplitude, the skyrmions tend to transiently adopt a quasi-coaxial configuration, which limits further increase in $v_\text{sp}$. While the excitation amplitude required to reach this threshold depends on $\alpha$, the threshold itself appears largely insensitive to damping.

\section{Orientational dynamics under thermal noise}
\label{sec.GilbertDamping}

As discussed in the main text, thermal noise introduces rich dynamics in self-propelled skyrmion pairs, with two main effects: (i) angular diffusion of the pair's heading direction, and (ii) spontaneous propulsion reversals. These noise-induced phenomena allow the pair to explore its environment, albeit through qualitatively distinct mechanisms. Here, we present additional data to help the reader distinguish between orientational diffusion and reversal events. First, we define the heading direction $\hat{\bm{n}}$ of the skyrmion pair as the unit vector perpendicular to the bond vector $\bm{\ell} \equiv \bm{R}_1 - \bm{R}_2$, namely, $\hat{\bm{n}} = \bm{\ell} \times \hat{\bm{z}} / \ell$, with orientation angle $\phi$. This should not be confused with the direction of motion, given by the instantaneous velocity $\bm{v}$ of the pair, which defines an angle $\theta$.

\begin{figure}[t]
\includegraphics[width=\linewidth]{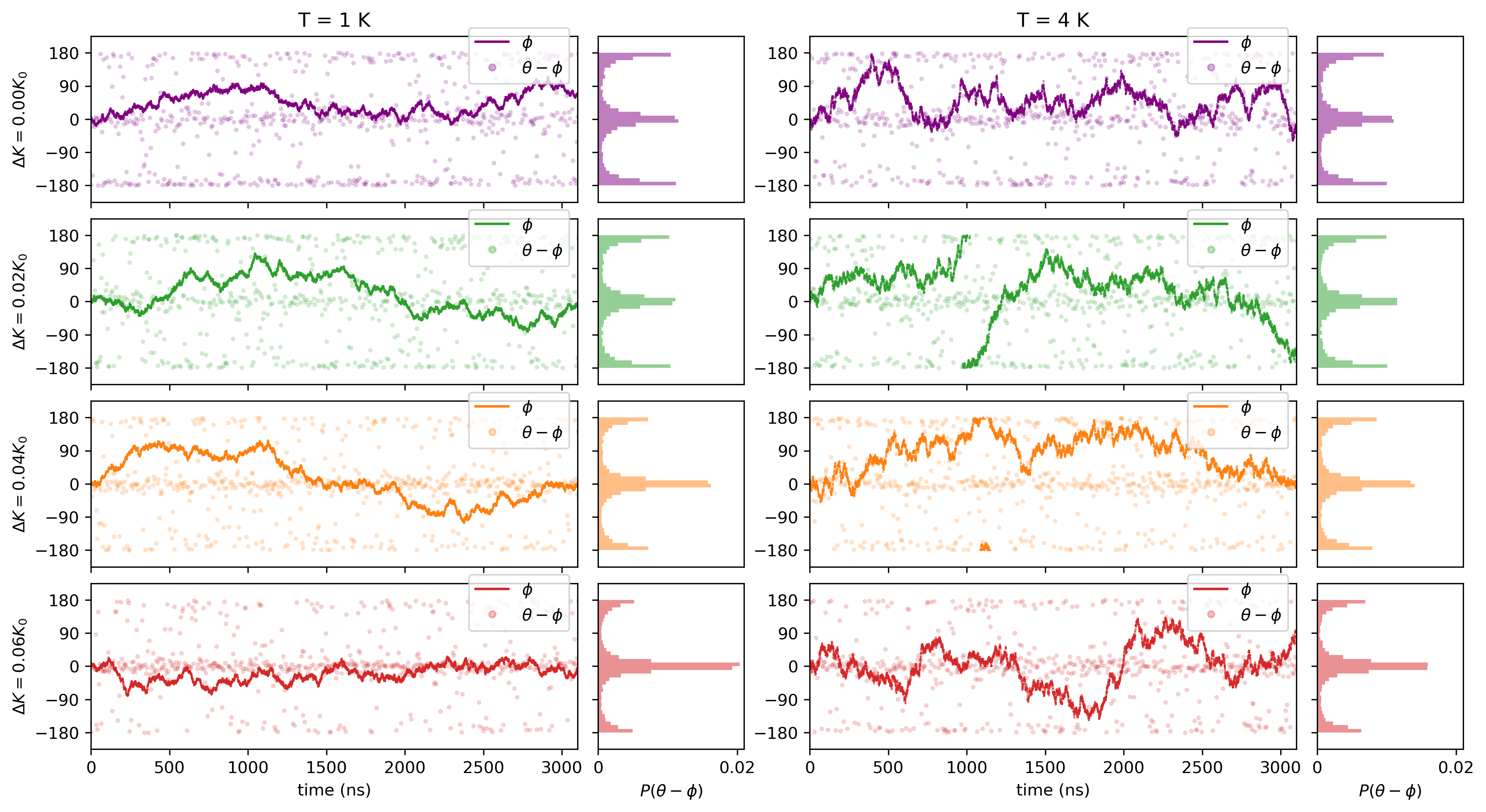}
\caption{Time series of the heading orientation $\phi$ of the skyrmion pair with respect to the $x$ axis in the trajectories shown in Fig. 4(a) and Fig. 4(b) of the main text, sampled every 0.5 ns. Here, the heading is defined as the direction normal to the bond, $\hat{\bm{n}}=\bm{\ell}\times\hat{\bm{z}}/\ell$. In all cases, the skyrmions were initialized heading east, that is, $\phi=0$. The symbols represent the orientation $\theta$ of the pair velocity, $\bm{v}$, relative to $\hat{\bm{n}}$. For a better visualization, the velocity data were sampled every 5 ns. The histograms depict the probability density function of $\theta-\phi$ for each time series. All angles are given in degrees and reduced to the interval $[-180^\circ,180^\circ]$.
}
\label{fig.ThetaPhi}
\end{figure}

Figure~\eqref{fig.ThetaPhi} shows time series of the heading angle $\phi$ for all trajectories presented in Fig.~4 of the main text. We also display the time dependence and corresponding histograms of the relative angle $\theta - \phi$ between the velocity vector and the heading direction. The histograms exhibit prominent peaks at $\theta - \phi = 0^\circ$, corresponding to forward motion, and at $\theta - \phi = \pm 180^\circ$, corresponding to propulsion reversal. At low anisotropy amplitudes ($\Delta K / K_0 \leq 0.02$), forward and backward motion occur with comparable probability, and the pair undergoes an effective one-dimensional random walk along the direction perpendicular to the bond. In contrast, at higher anisotropy amplitudes ($\Delta K / K_0 \geq 0.04$), forward motion dominates, leading to strongly persistent motion along the heading direction. As expected, the probability peak at the backward direction becomes stronger for higher temperatures, indicating an increase in the rate of reversals.

\end{document}